\documentclass[preprint,12pt]{elsarticle}

\usepackage{amssymb}
\journal{Annals of Physics}

\begin{document}

\begin{frontmatter}



\title{An improved derivation of minimum information quantum gravity}


\author{P. A. Mandrin}

\address{Department of Physics, University of Zurich, Winterthurerstrasse 190, 8057 
Z\"urich, CH. E-mail: pierre.mandrin@uzh.ch.}

\begin{abstract}
Minimum information quantum gravity (MIQG) is a theory of quantum gravity which requires no explicit microscopic quantum structure. In this article, it is shown that the MIQG action can be derived using a more elegant and straight-forward method than in the first existence proof. The required assumptions are dramatically reduced. In particular, former assumptions referring to the existence of quantum boxes, the exact differential of the entropy variation and the role of the boundary can be omitted. Moreover, the open problem of the quantum occupation number per box is solved. Thus, the arguments in favour of MIQG become even more stringent. The remaining assumptions are 1. the principle of optimisation of the resulting per imposed degrees of freedom, 2. abstract quantum number conservation, 3. the validity of the laws of thermodynamics, 4. identification of a macroscopic parameterisation with space-time and 5. unspecific interactions. Although the requirements are reduced, all former results remain valid. In particular, all well established physics as special cases (Quantum Field Theory, QFT, and General Relativity, GR) follow and all measurable quantities may be computed.
\end{abstract}

\begin{keyword}
Quantum Gravity \sep Entropy \sep General Relativity \sep Quantum Mechanics



\end{keyword}

\end{frontmatter}



%%%%%%%%% 1. Introduction

\section{Introduction}
\label{intro}

Many theories of quantum gravity exist, but they are mostly facing major conceptual difficulties (see e.g. \cite{Carlip}). A common feature is that they assume a priori quantum dynamics and thus require quantisation of initially classic quantities. However, no consensus exists on what should be the starting point for quantisation. E.g. explicitly covariant quantisation might be required as proposed in \cite{Kanatchikov}, while loop quantum gravity allows fewer momentum components (see also argumentations in  \cite{Thiemann}). As an alternative, several attempts have been made to let space-time and gravity emerge \cite{Sindoni}. They usually impose assumptions like the intrinsic space-time structure or the ''holographic'' nature of boundaries, as e.g. in \cite{Gogberashvili}. In emergent scenarios, hardly any more explanation is gained on the ''unknown'' properties of the quanta.

\vspace{5mm}
All these difficulties could be avoided by starting with less assumptions at the beginning, as proposed in MIQG\footnote{Minimum information quantum gravity.}. MIQG is a theory based on abstract quantum number conservation, the laws of thermodynamics, unspecific interactions, and locally maximises the ratio of resulting degrees of freedom per imposed degree of freedom of the theory. No explicit dynamical structure is required on the microscopic level (no Lagrangian or Hamiltonian description). According to MIQG, quantum measurements are interpreted as measurements performed macroscopically on the quantum detectors themselves.

\vspace{3mm}
The concept of MIQG has been introduced in \cite{Mandrin1} and \cite{Mandrin2}. However, some initial assumptions related to pre-MIQG physics are not necessary, and the derivation of the general results and the special cases of GR and QFT can be performed in a more systematic and comprehensive way. The improved derivation is performed in this article, and additional information is gained on the model.

\vspace{3mm}
This article shows that the action can be derived:

\begin{enumerate}%[1.]
\item without needing to introduce a priori quantum boxes out of any context,
\item without needing to identify the variation integrand of the entropy with an exact differential of the entropy density, and 
\item without needing to postulate (a priori)  the boundary contribution of the entropy as an expression describing the volume enclosed.
\end{enumerate}

\vspace{3mm}
Nevertheless, all results of \cite{Mandrin1} and \cite{Mandrin2} remain valid, including the results concerning the generalised gravitational action and concerning quantum mechanical behaviour and the quantisation prescription.

\vspace{3mm}
The new procedure resides in identifying the number of boxes comprised between given (boundary) locations with the entropy induced by the space-time parameterisation itself. It then suffices to use the vielbeins as the only varying gravitational parameters. This method automatically leads to the boundary expression for the entropy, from which all former results follow.

\vspace{3mm}
After presentation of the new derivation of the action, it is shown how quantum gravity evaluations may be performed properly on the basis of suitable quantum detectors, and the open problem of maximum quantum occupation numbers is clarified.

%%%%%%%%% 2. Derivation of the entropy conformly to the new concept

\section{Derivation of the entropy conformly to the new concept}
\label{sec:2}

% 2.1
\subsection{Quanta, parameterisation and boxes}

Let us introduce first one species of quanta and call them ''primary quanta''. In order to use thermodynamics, we need to consider statistically large numbers of these quanta. Thermodynamically well behaved systems of quanta (i.e. statistically large and in equilibrium) allow us to define macroscopic quantities. Such systems are called macroscopically separable systems (according to \cite{Mandrin1} or appendix A). 

\vspace{3mm}
Conformly to information theory, a system $\mathcal{S}$ is in thermal equilibrium if its macroscopic state maximises the number of possible microscopic states. This condition is also satisfied if, for an arbitrary partition of  $\mathcal{S}$ into many subsystems, the number of possible distributions of quanta among the subsystems is maximised. The latter (stronger) condition is used for MIQG (see appendix A).

\vspace{3mm}
In the context of black holes, space-time has been shown to have the properties of macroscopic variables (interpretation of the horizon area as an entropy \cite{Bekenstein} and of the surface gravity as a temperature causing radiation \cite{Hawking}). The outer vicinity of the black hole horizon should be considered as a special case of a more general physical theory. Moreover, the variation of statistical and Wald's entropy have been found to be equivalent \cite{Hadad_Kupferman}. Yet the nature of macroscopic and microscopic quantities are fundamentally incompatible. For this reason, in MIQG, space-time is postulated to be defined by macroscopic variables.

\vspace{3mm}
We always may define in a completely arbitrary way an ordering of the quanta and thus of the systems. This gives a two-fold structure. For a given system $\mathcal{S}$, we may \\
\begin{enumerate}
\item define adjacent systems $\mathcal{S}_{left} <: \mathcal{S}$ ($\mathcal{S}_{left}$ has the covering $\mathcal{S}$) and $\mathcal{S}_{right} :> \mathcal{S}$ and \\
\item attach to the left and right ''ends'' of $\mathcal{S}$ distinct but otherwise arbitrary values of a macroscopic parameter $x$.
\end{enumerate}

\vspace{3mm}
By construction, this parameterisation is macroscopic. Because any macroscopically separable systems may be partitioned into macroscopically separable subsystems, the parameterisation is also dense. On the other hand, systems with small numbers of quanta (we call such systems boxes) are not dense. The quanta are still ordered, but no parameterisation is defined. We can merely partition a macroscopically separable system $\mathcal{S}$ into many boxes. We do not have any constraint on how ''small'' a box should be. Therefore, the maximum box occupation number $p$ (number of quanta fitting into each box $+1$, $p \in \mathbb{N}$) can be fixed arbitrarily before starting the computations. Without loss of generality, we may choose the same number $p$ for each box, as a convention. This is how the notion of boxes emerges.

% 2.2
\subsection{Thermodynamics, entropy and Lorentzian space}

Consider a system $\mathcal{S}$ of $N$ quanta distributed in $n_S$ boxes. We introduce a quantity $E = e_0 \ N$, where $e_0$ is a scaling constant and plays a similar role as $\hbar$, {\rm i. e.} $e_0$ tells us how much of $E$ is carried per quantum object. An infinitesimal change of entropy $S$, without changing the distribution of quanta inside $\mathcal{S}$, is given by 

\begin{equation}
\label{entropy}
\delta S = \ln{p} \ \delta n_S .
\end{equation}

\noindent Therefore, the following equalities trivially hold:

\begin{eqnarray}
\label{eq:numbers}
\delta N & = & \rho_S \ \delta n_S, \\
\label{eq:First}
\delta E & = & T \ \delta S,
\end{eqnarray}

\noindent where $\rho_S$ is the mean quantum occupation number per box, and $T \sim \rho_S$ is the temperature (or ''mean box filling indicator''). According to the thermal equilibrium condition, any macroscopically separable subsystem $\mathcal{S}_l$ of $\mathcal{S}$ has the same temperature, i.e. $T_l = T$. Eqn. (\ref{eq:First}) is the First Law of thermodynamics in its simplest form.

\vspace{3mm}
It can be shown that a space $\mathcal{M}$ of smooth macroscopic parameterisation may locally be represented by a vector space $V$ isomorphic to $\mathbb{R}^n$ with $n \in \mathbb{N}$, and there is a locally trivial fiber bundle $E$ over $\mathcal{M}$ with standard fiber $V$ (see Appendix A).

\vspace{3mm}
In addition, the value of the dimension $n$ and the detailed structure of $V$ may be determined. Conformly to \cite{Mandrin1} and especially appendix B (for a more complete treatment), $n = 3+1$ precisely optimises the ratio of locally resulting degrees of freedom per imposed degree of freedom. It follows that $\mathcal{M}$ locally has the structure of Lorentzian space. Moreover, the physical laws must be diffeomorphism invariant in the approximation of thermal equilibrium, and the space of diffeomorphisms of Lorentzian space is known to be spanned by the tetrads $e^I_\mu$ as a convenient orthonormal basis.

\vspace{3mm}
Thermal equilibrium may fail to apply to arbitrary large $\mathcal{M}$. Thus, it is necessary to restrict Eqns (\ref{eq:numbers}) and (\ref{eq:First}) to thermally small $\mathcal{M}$, i.e. in approximate thermal equilibrium (see \cite{Mandrin1} or Appendix A). 
The entropy of a thermally small region $\mathcal{M}$ may be approximated as a sum over the entropies $S_k$ of macroscopically separable systems $\mathcal{S}_k$ contained in $\mathcal{M}$, with $k=1,\ldots,m$:

\begin{equation}
\label{eq:S_sum_M}
S = \sum_{k=1}^{m} S_k.
\end{equation}

\noindent We may obtain an accurate expression by requiring the systems $\mathcal{S}_k$ to be infinitesimal. The sum is then converted into an integral with integrand $\lambda$, the entropy 4-density:

\begin{equation}
\label{eq:S_int_M_s}
S = \int_\mathcal{M} {\rm d}^4x \ \lambda.
\end{equation}

Because the number of boxes of a system is not affected by a change of its parameterisation, $S$ behaves as a scalar. However, $\lambda$ is not a scalar. In order to recover a scalar, we introduce the metric $g_{\mu\nu} = e^I_\mu \ \eta_{IJ} \ e^J_\nu$, with determinant $g$. Then, decompose $\lambda = \sqrt{-g} \ \mathcal{L}$, so that $\mathcal{L}$ is a scalar and is the entropy 4-density in Minkowski space representation, and ${\rm d}^4x \ \sqrt{-g}$ is the 4-volume element in Minkowski space representation:

\begin{equation}
\label{eq:S_int_M}
S = \int_\mathcal{M} {\rm d}^4x \ \sqrt{-g} \ \mathcal{L}.
\end{equation}

%2.3
\subsection{Second Law of thermodynamics and gravity}

We have to apply the Second Law of thermodynamics $\delta S = 0$ to $\mathcal{M}$. Thermal smallness allows to simplify the computation by starting from a boundary integral as shall be shown next. Consider changes of the value of $S$. These are due either to modifications of the quanta contained in $\mathcal{M}$ or to modifications of the shape of the boundary $\partial \mathcal{M}$ of $\mathcal{M}$. However, changing the shape of the boundary is equivalent to adding or removing part of the bulk quanta, which we can see by integrating by parts as follows. Without loss of generality, choose $\mathcal{M}$ together with a gauge $(x^\mu)$ such that $g_{\alpha\beta} = 0$ and $g_{\beta\gamma} \approx \eta_{\beta\gamma}$ for a given $\alpha$ and for any $\beta, \gamma \ne \alpha$, $\alpha \in \{0, 1, 2, 3, 4\}$, and such that two opposite boundary components, say $\Sigma_{\alpha+}$ and $\Sigma_{\alpha-}$,  are parameterised without dependence on the component $x^\alpha$. First consider the case for which the other boundaries have much smaller ''hypersurface areas'' than $\Sigma_{\alpha+}$ and $\Sigma_{\alpha-}$ (thin layer). Then, the integral over $\mathcal{M}$ reduces to

\begin{eqnarray}
\delta S & = & \delta \int_\mathcal{M} {\rm d}^4x \ \sqrt{-g} \ \mathcal{L} \nonumber \\
\label{eq:boundary}
& \approx & \delta [\int_{\Sigma_{\alpha+}} {\rm d}^3x \ \sqrt{|g_{\alpha\alpha}|} \ \sqrt{|\gamma|} \ s_\alpha - \int_{\Sigma_{\alpha-}} {\rm d}^3x \ \sqrt{|g_{\alpha\alpha}|} \ \sqrt{|\gamma|} \ s_\alpha],
\end{eqnarray}

\noindent where $\gamma$ is the determinant of $g_{\mu\nu}$ after removing the $\alpha$th line and row, $s_\alpha$ is the integral function of $\mathcal{L}$ with respect to $x^\alpha$, integration by parts has been applied with respect to $x^\alpha$, yielding a boundary term, and the second (bulk) term vanishes because the fixed 3d-averaged values of $\sqrt{|g_{\alpha\alpha}|}$, written as $\overline{\sqrt{|g_{\alpha\alpha}|}}$, have vanishing derivative:

\begin{eqnarray}
\nabla_\alpha \overline{\sqrt{|g_{\alpha\alpha}|}} & = & \nabla_\alpha \frac{\delta S}{\delta [\int_{\Sigma_{\alpha+}} - \int_{\Sigma_{\alpha-}}] \ {\rm d}^3x \ \sqrt{|\gamma|} \ s_\alpha}
 \nonumber \\
\label{eq:nabla_gaa}
& \approx & \nabla_\alpha \frac{\delta S}{\delta \int_{\mathcal{M}} {\rm d}^4x \ \sqrt{|\gamma|} \ \mathcal{L}} = \nabla_\alpha \frac{\delta S}{\delta N} \sim \nabla_\alpha T^{-1} = 0, 
\end{eqnarray}

\noindent by the thermal equilibrium condition. In other words, the structure of the quanta inside $\mathcal{M}$ is irrelevant because of thermal equilibrium. This result may easily be generalised to arbitrary thermally small $\mathcal{M}$, by patching together many thin layers with all 4 orientations. Thus, in order to recover the total variation $\delta S_{\rm total}$, it is sufficient to evaluate the boundary term of $\delta S$ as claimed above. 

\vspace{3mm}
For arbitrary gauge and arbitrary thermally small $\mathcal{M}$, the procedure for obtaining $\delta S_{\rm total}$ must be to vary the boundary term with respect to the quantum density contribution (the boundary shape fixes the gauge) and then to add a bulk integral term of the same form but varied with respect to the gauge $e^I_\mu$ (in order to obtain the total variation of an expression of the form of Eqn. (\ref{eq:S_int_M})). Interestingly, this procedure happens to coincide with the standard variation procedure used in \cite{Mandrin1} and \cite{Mandrin2} as well as in \cite{York} for the analogous variation of the action.

\vspace{3mm}
Consider now the variation of the integrand of the boundary contribution to Eqn. (\ref{eq:S_int_M}). Because the parameterisation induces the entropy, we vary $S$ as a function of the triads $e^I_i$, where lower-case latin indices refer to the 3-dimensional subspace $\tilde{V}$ of $V$ associated to the boundary at a given boundary point $p$. The variation of the entropy must be a scalar. These conditions are fulfilled by

\begin{eqnarray}
\delta S\bigg|_{\partial\mathcal{M}} & = & \sum_{a} \int_{\partial\mathcal{M}_a} {\rm d}^3x \ \sqrt{|g_{\perp\perp}|} \ \sqrt{|\gamma|} \ f^i_{a I} \ \delta e^I_i \nonumber \\
\label{eq:var_boundary_triads}
& = &  \sum_{a} \int_{\partial\mathcal{M}_a} {\rm d}^3x \ \sqrt{|\gamma|} \ \tau^i_{a I} \ \delta e^I_i.
\end{eqnarray}

\noindent where, without loss of generality, the boundary is smooth everywhere except on the 2-dimensional intersections between different boundary components $\partial\mathcal{M}_a$ ($a=1, \ldots, a_{max}$) which are chosen to be normal to each other (the scalar product being well-defined), $\gamma_{ij}$ and $n^\mu = \delta_\perp^\mu$ denote the intrinsic metric and the ''normal vector'' with respect to the submanifold $\partial\mathcal{M}_a$ (as defined by the respective subspaces)\footnote{Subspaces normal to each other are well-defined by means of the scalar product defined in $V$. However, neither can we assume $V$ to be tangent to the manifold, nor can we assume the subspace $\tilde{V}$ to be tangent to $\partial\mathcal{M}_a$.}, and the not yet specified coefficients $f^i_{a I}$ or $\tau^i_{a I} = \sqrt{|g_{\perp\perp}|} \ f^i_{a I}$ must be 3-vector-valued functions of the location on the boundary. 

\vspace{3mm}
Incidently, Eqn. (\ref{eq:var_boundary_triads}) corresponds to the resulting boundary term of \cite{Mandrin2} and also is analogous to the boundary term of the quasi-local action as found in \cite{Brown_York_1992}, \cite{Brown_York_1993} and \cite{Creighton_Mann}. Even then, the quantity $\tau^i_{a I}$ cannot be the same as the analogous quantity for GR, as has already been shown in \cite{Mandrin2}. To be prudent, we should give $\tau^i_{a I}$ at most the surname of analogous surface stress density of MIQG. Eqn. (\ref{eq:var_boundary_triads}) has been obtained without needing any considerations about exactness of any variational differentials, as compared with \cite{Mandrin1}, i.e. no pre-MIQG assumptions about physics have been required. This means that MIQG does not leave any open door for ambiguities of any kind during all the computations.

\vspace{3mm}
In order to obtain the desired variation with respect to the quantum density, and not with respect to the gauge, one has to perform a Legendre transformation. We may also argue in another way \cite{Mandrin2}. Changes of $e^I_i$ affect the volume of $\mathcal{M}$ and thus also the volume of neighbouring space-time regions. Therefore, $\mathcal{M}$ must lie in a ''bath of tetrads'', requiring a Legendre transformation. We thus obtain:

\begin{equation}
\label{eq:var_boundary_tau}
\delta S\bigg|_{\partial\mathcal{M}} = \sum_{a} \int_{\partial\mathcal{M}_a} {\rm d}^3x \ \sqrt{|\gamma|} \ e^I_i \ \delta \tau^i_{a I}.
\end{equation}

\vspace{3mm}
Using Eqn. (\ref{eq:var_boundary_tau}) and the procedure described above, the entropy without interactions is obtained as computed in \cite{Mandrin2} or Appendix C:

\begin{equation}
\label{eq:Total_grav}
S_{\rm total} = \int_\mathcal{M} d^4x \ \sqrt{-g} \ [e^I_\mu \ e^J_\nu \ \Phi^{\mu\nu}_{IJ} + \omega_{\mu IJ} \ \Omega^{\mu IJ}]
\end{equation}

\vspace{3mm}
Eqn. (\ref{eq:Total_grav}) exhibits the simultaneous dependence of the entropy on the tetrads $e^I_\mu$ and the connection 1-form $\omega_{\mu IJ}$. For vanishing torsion and if we expand $\Phi^{\mu\nu}_{IJ}$ with respect to dimensions of increasing order in the derivative, the lowest order entropy is in one-to-one correspondence with the Palatini action describing the gravitational field of GR,

\begin{equation}
\label{eq:Palatini}
S_{\rm Palatini} = \int_\mathcal{M} d^4x \ \sqrt{-g} \ [e^I_\mu \ e^J_\nu \ F^{\mu\nu}_{IJ} + \Lambda],
\end{equation}

\noindent with the cosmological constant $\Lambda$. We can therefore identify $\Phi^{\mu\nu}_{IJ}$, to lowest order, with the curvature 2-form $F^{\mu\nu}_{IJ}$ plus $\Lambda \ e_I^\mu \ e_J^\nu$. Furthermore, $e^I_\mu$ may be interpreted to be the gravitational field and $\mathcal{M}$ to be (a piece of) generalised space-time manifold defined by $e^I_\mu$ and $\omega_{\mu IJ}$.

% 2.4
\subsection{Angular momentum}
\label{sec:ang_mom}

It is straightforward to derive the ADM-decomposition of Eqn. (\ref{eq:var_boundary_tau}), following the same procedure as in \cite{Brown_York_1992} or \cite{Brown_York_1993} and inserting the triad notation of \cite{Mandrin2}. The projections are performed onto subspaces of $V$. The Euclidean subspace shall be called space-like, the 1d-subspace orthogonal to it shall be called time space (even if not tangent), and the 3d-subspaces containing the time-space shall be called time-like. The boundaries shall be named correspondingly. Consider the term from the time-like boundary ($\mathcal{T}$) and perform one more Legendre transformation $s^i_I \leftrightarrow e^I_i$, conformly to permeability with respect to stress \cite{Mandrin2}:

\begin{equation}
\label{eq:var_boundary_ADM}
\delta S\bigg|_\mathcal{T} = \int_\mathcal{T} {\rm d}^3x [ N \delta ({\sqrt{\sigma} \epsilon}) - N^i \delta (\sqrt{\sigma} j_i) + N \sqrt{\sigma} s^i_I \delta e^I_i].
\end{equation}

\noindent with generalised lapse $N$, shift $N^i$, surface energy density $\epsilon$, surface momentum density $j_i$, and stress vector $s^i_I$. Consider $\mathcal{M}$ in local Minkowski coordinates. Approximate time translation and rotation isometries allow an extension of the First Law. The corresponding conserved quantities are the ''inner energy'' $U$ and the angular momentum $\mathcal{J}$, respectively, which remain constant if no exchange with the outer space occurs. Eqns (\ref{eq:var_boundary_tau}) and (\ref{eq:var_boundary_ADM}) tell us that the First Law of thermodynamics for $\mathcal{M}$ is of the following form (see also \cite{Mandrin1}):

\begin{equation}
\label{eq:first_lawJP}
\delta S = T^{-1} \ (\delta U - \omega^\mu \ \delta\mathcal{J}_{\mu} + P_I^\mu \ \delta V_\mu^I).
\end{equation}

\noindent The symbols appearing in Eqn. (\ref{eq:first_lawJP}) are explained in \cite{Mandrin1} and partly rewritten in triad notation. The second term on the right-hand-side is generated by rotations which leave $S$ invariant. We may keep $U$ and $V_\mu^I$ fixed at their physical values and vary $\mathcal{J}_\mu$ alone:

\begin{equation}
\label{eq:first_lawJ}
\delta S = \varsigma^\mu \ \delta\mathcal{J}_{\mu},
\end{equation}

\noindent where $\varsigma^\mu = T^{-1} \ \omega^\mu$. Eqn. (\ref{eq:first_lawJ}) means that entropy is induced by arbitrarily distributing secondary quanta, namely the quanta of angular momentum. 
One could just as well use the angular momentum instead of $E$ in order to parameterise space-time regions on which the angular momentum density does not vanish. Therefore, the concept of arbitrarily fixed maximum box occupation number $p$ may be transferred to the angular momentum quantum number.

% 2.5
\subsection{Matter}

The matter contribution to the total action is obtained by allowing for exchange of quanta between physical systems. The procedure is shown in  \cite{Mandrin1} and  \cite{Mandrin2}. The action is derived starting from the following interaction boundary term (here in the example of the $\mathcal{T}$-term):

\begin{equation}
\label{eq:var_matter}
\delta S_{\rm matter}\bigg|_{\mathcal{T}} = \sum_A \int_\mathcal{T} {\rm d}^3x \ \sqrt{-\gamma} \ \Pi^{IA} \ \delta A_{IA},
\end{equation}

\noindent where $A_{IA}$ is the interaction potential with indices $I = i_1, i_2, \ldots i_q$, $q \in \mathbb{N}_0$, for the type $A$ interaction and $\Pi^{IA}$ is the conjugate potential. The same procedure as in Subsection \ref{sec:ang_mom} leads to the total action \cite{Mandrin2}%%

\begin{equation}
\label{eq:action_final}
S_{\rm total} = \int_{\mathcal{M}}  {\rm d}^4x \ \sqrt{-g} \ [e^I_\mu \ e^J_\nu \ \Phi^{\mu\nu}_{IJ} + \omega_{\mu IJ} \ \Omega^{\mu IJ} + \sum_A j'^{\Gamma A} \ A_{\Gamma A} + F'^{\Delta A} \ F_{\Delta A}],
\end{equation}

\noindent where the field $F^{\Delta A}$ is the (possibly anti-symmetrised) covariant derivative of $A^{\Gamma A}$ with indices $\Gamma = i_0, i_1, \ldots i_q$ and $\Delta = i_0, i_1, \ldots i_{q+1}$, and $j'^{\Gamma A}$ and $F'^{\Delta A}$ are the generalised current 4-density and generalised field, respectively, as defined in \cite{Mandrin2}.

\vspace{3mm}
Eqn. (\ref{eq:var_matter}) also has a corresponding term in the First Law, with chemical potential $\mu^A$ and particle number $n_{pA}$, as follows \cite{Mandrin1}:

\begin{equation}
\label{eq:first_lawJPI}
\delta S = T^{-1}(\delta U - \omega^\mu\delta\mathcal{J}_{\mu} + P_I^\mu \delta V_\mu^I + \mu^A \delta n_{pA}).
\end{equation}

Because the charges are conserved quantities, it is possible to introduce a particle number quantum number in the same manner as for the angular momentum. Accordingly, the maximum occupation number may be fixed to an arbitrary value. In the case of bosonic matter, there is hence no restriction on how many particles are allowed to be in the same quantum state. For fermions, one needs to exploit the derivation of QFT according to \cite{Mandrin1}. For this reason, the QFT-like formulation of quantum behaviour is readdressed shortly in the following section.

%%%%%%%%% 3. Quantum measurements

\section{Quantum measurements}
\label{sec:3}

In MIQG, although quantum space-time dynamics are not provided, quantum measurements can still be described. This is because quantum measurements are interpreted as being performed on the quantum detector itself, which is a macroscopic system in unstable thermal equilibrium \cite{Mandrin1}. The reader is also refered to \cite{Mandrin1} regarding the interpretation (from this perspective) of wave function collapse, unitarity and the loss of memory of quantum states by subsequent measurements (e.g. Stern-Gerlach-type experiments).

\vspace{3mm}
It is also possible to recover the second quantisation method for matter fields (which yields QFT) under the condition of negligible gravitational field, as shown in \cite{Mandrin1}. The condition of negligible gravitational field is necessary because the standard quantisation method simulates absorption and creation of one field particle which must be described in the same positive energy mode both at creation and at detection (i.e. non-locally). Typically, the number $n_A(k)$ of $A$-type particles in $k$-mode is replaced by the operator $\hat{a}_k^\dagger \hat{a}_k$. 

\vspace{3mm}
Even in presence of significant gravity, quantum measurements can be transferred to the macroscopic domain, thus the result of measurements can still be predicted using QFT methods provided the detector is parameterised in local Minkowski coordinates and the detector satisfies the condition of thermal smallness in these coordinates.

\vspace{3mm}
Consider particles of interaction type $A$ described by the potential function $A_{IA}$ and with compact support given by $\mathcal{M}$.
Suppose that the efficiency of detecting such a particle that just ''crosses'' the trajectory of the detector is equal to $\eta_A$ and that the detector has the 4-volume $\mathcal{M}_D$.  Then, the probability that a particle is detected is given by

\begin{equation}
\label{eq:q_meas}
p_D = \eta_A \frac{Q_A\big|_{\mathcal{M}_D}}{Q_A\big|_\mathcal{M}} = \eta_A \frac{\int_{\mathcal{M}_D} {\rm d}^4x \ \sqrt{-g} \ j'^{0i_2\ldots,A}(x)}{\int_{\mathcal{M}} {\rm d}^4x \ \sqrt{-g} \ j'^{0i_2\ldots,A}(x)},
\end{equation}

\noindent where $Q_A$ is the type $A$ charge \cite{Mandrin2} due to many identically prepared type $A$ particles in the macroscopic field (ignoring the detector type $A$ charge contribution).

\vspace{3mm}
On the other hand, QFT predicts the detection probability

\begin{eqnarray}
p_D & = & \eta_A \frac{\int_{\Sigma_D} {\rm d}^3x \ \langle \vec{x} | \hat{N}_A | \vec{x} \rangle}{\int_\Sigma {\rm d}^3x \ \langle  \vec{x} | \hat{N}_A | \vec{x} \rangle} \nonumber \\
& = & \eta_A \frac{\int_{\Sigma_D} {\rm d}^3x \ \int {\rm d}^3p \ [(2\pi)^3 \ 2p_0(|\vec{p}|)]^{-1} \ \langle \vec{x} | \vec{p} \rangle \ \langle \vec{p} | \hat{a}^\dagger \ \hat{a} | \vec{p} \rangle \ \langle \vec{p} | \vec{x} \rangle}{\int_\Sigma {\rm d}^3x \ \int {\rm d}^3p \ [(2\pi)^3 \ 2p_0(|\vec{p}|)]^{-1} \ \langle \vec{x} | \vec{p} \rangle \ \langle \vec{p} | \hat{a}^\dagger \ \hat{a} | \vec{p} \rangle \ \langle \vec{p} | \vec{x} \rangle} \nonumber \\
& = & \eta_A \frac{\int_{\Sigma_D} {\rm d}^3x \ \int {\rm d}^3p \ [(2\pi)^3 \ 2p_0(|\vec{p}|)]^{-1} \ \langle \vec{x} | \vec{p} \rangle \ \langle \vec{p} |  \hat{Q}_A | \vec{p} \rangle \ \langle \vec{p} | \vec{x} \rangle}{\int_\Sigma {\rm d}^3x \ \int {\rm d}^3p \ [(2\pi)^3 \ 2p_0(|\vec{p}|)]^{-1} \ \langle \vec{x} | \vec{p} \rangle \ \langle \vec{p} | \hat{Q}_A | \vec{p} \rangle \ \langle \vec{p} | \vec{x} \rangle} \nonumber \\
\label{eq:QFT_meas}
& = & \eta_A \frac{\int_{\mathcal{M}_D} {\rm d}^4x \ j'^{0i_2\ldots,A}(x)}{\int_\mathcal{M} {\rm d}^4x \ j'^{0i_2\ldots,A}(x)},
\end{eqnarray}

\noindent where $\Sigma_D$ is the 3-volume of the detector, $\Sigma$ is the compact 3d-support of the particle wave function, $\hat{N}_A$ is the particle number operator, $\hat{Q}_A$ is the charge operator and $| \vec{x} \rangle$ is a state in position representation, and the 3d-integral has been converted to a 4d-integral while replacing (the expectation value of) the 3-density $q_A|\langle \vec{p} | \vec{x} \rangle|^2$ by a 4-density $j'^{0i_2\ldots,A}$ ($q_A$ being the type $A$ charge of the particle). Eqn. (\ref{eq:QFT_meas}) corresponds to Eqn. (\ref{eq:q_meas}) in the limit of flat space ($g=-1$).

\vspace{3mm}
Bosonic and fermionic matter differ in respect to the form of the Hamiltonian, expressed for negligible gravitational field. Following the standard QFT approach, bosonic matter requires the creation and absorption operators to satisfy the commutation relation $[\hat{a}_k,\hat{a}^\dagger_{k'}]\sim \delta(k-k')$ in order to satisfy the principle of causality, whereas fermionic matter requires the creation and absorption operators to satisfy the anticommutation relation in order for the energy of particles and antiparticles to be bounded from below.

It follows that bosons keep the arbitrary particle occupation number condition as derived above, whereas fermions are subject to an additional constraint due to Fermi-Dirac-statistics, thus allowing only one particle to occupy the same state. This means that MIQG is compatible with many-particle statistics of QFT.

%%%%%%%%%%%%% 4. Conclusions

\section{Conclusions}
\label{sec:4}

In this article, it has been shown that MIQG can be constructed without requiring any assumptions of any former physical theory or postulate, except for unspecified interactions. The number of necessary assumptions has been significantly reduced compared to the first existence proof of MIQG, and the arguments supporting this approach are even stronger. This is because this approach has no ambiguity of any kind concerning the derivation of the quantum gravity results. Nevertheless, all the former results remain valid, and all established physical phenomena are reproduced as special cases. All measurable quantities are computable, including the probability of detecting a given particle in the strong gravitational field, if a suitable detector is chosen. Finally, the quantum boxes are a byproduct of macroscopic space-time parameterisation, and the open problem of the maximum occupation number per box is resolved and, in particular, is in accord with Einstein-Bose-statistics and Fermi-Dirac-statistics.

%%%%%%%%%%%%%%%% Acknoledgements

\section*{Acknowledgements}
I would like to thank Philippe Jetzer for hospitality at University of Zurich.

%%%%%%%%%%%%%%    APPENDIX

\appendix

%%%%%%%%%%%%% Appendix A

\section{}
\label{sec:A}

In this appendix, a few mathematical tools for MIQG are summarised and proofs are given.

\vspace{3mm}
In MIQG, thermal equilibrium for systems with large numbers of quanta must be defined without imposing any quantum structure a priori. Therefore, we must consider all possible distributions of the quanta among arbitrary many subsystems.

\subsection{Definition: Thermal equilibrium}
A system $\mathcal{S}$ of quanta is in thermal equilibrium in the sense of MIQG if, for any partition of $\mathcal{S}$ into $N_s \gg 1$ subsystems with finite maximum quantum number for each subsystem, the number of possible distributions of the quanta among the subsystems is maximised.

\vspace{3mm}
This definition of thermal equilibrium is at least as strong as the definition according to Shannon's information theory, as can be seen for a model with explicit microstructure: 

\paragraph{Claim: Thermal equilibrium}
Be a model in which a system $\mathcal{S}$ contains a set of $n_L$ possible quantum ''locations''. Distribute $N$ quanta among them with at most $p_0-1$ quanta per location. Then, if the system is in thermal equilibrium in the sense of MIQG, it also maximises the number of possible microstates ({\rm i. e.} the number of possible distributions among the locations). 

\paragraph{Proof:}
Suppose that there existed a system $\mathcal{S}$ in thermal equilibrium in the sense of MIQG representing a macroscopic state which does not maximise the number of microstates. Consider a sequence $\{P_k=\{\{\mathcal{S}_{k,j}, j=1\ldots m_k\}, k \in \mathbb{N}\}$ of partitions $\{\mathcal{S}_{k,j}, j=1\ldots m_k\}$ of $\mathcal{S}$, with $m_{k+1} = 2m_k$ and $\{\mathcal{S}_{k+1,2j-1}, \mathcal{S}_{k+1,2j}\}$ is a partition of $\mathcal{S}_{k,j}$. For $k \rightarrow \infty$, the partitions would tend to a one-to-one representation of the macrostate. Because the macrostate would not maximise its number of microstates, there would exist another macrostate with more microstates which may be represented by another sequence $\{P'_k\}$ of partitions, so that $\exists k_0$ with $P'_k \ne P_k \forall k>k_0$. It is alway possible to find a common partition $P''_k$ of both $P_k$ and $P'_k$, so that the numbers of possible microstates for $P^*_k = P_k, P'_k$ are given by (or, for $k \rightarrow \infty$, tend to)

\begin{equation}
\label{eq:microstates}
\Omega^* = \sum_{j=1}^{m_k} p_0^{n^*_L}
\end{equation}

\noindent (the star stands for either prime or not prime). Thus, if $\Omega' > \Omega$, then $n'_L > n_L$ for some $j$, and this can only be achieved if the number of subsystems $n'_s > n_s$ for the same $j$. Therefore, $P_k$ does not maximise the number of possible distributions of quanta among the $\mathcal{S}''_{k,j}$. This is in contradiction to the thermal equilibrium condition of $\mathcal{S}$ in the sense of MIQG. Therefore, the claim is proven.

\subsection{Definition: Macroscopically separable sytems}
The systems $\mathcal{S}_k$ of quanta in a set $\{\mathcal{S}_k, k=1\ldots m\}$ are macroscopically separable if
\begin{enumerate}
\item a statistically large number $n_k$ of quanta may be assigned to each system $S_k$, \mbox{i. e.} the statistical fluctuations of quanta associated to each system are negligible, $n_k \gg \sqrt{n_k}$,\\
\item each system is in thermodynamic equilibrium (in the sense of MIQG),\\
\item for each pair of systems $\mathcal{S}_i$, $\mathcal{S}_j$, there exists at least one well-defined thermodynamic (macroscopic) variable in respect to which they differ. Normally, this variable will be the location.
\end{enumerate}

\subsection{Definition: Smooth system}

A system $\mathcal{S}$ is smooth if, for any sequence $\{P_k=\{\{\mathcal{S}_{k,j}, j=1\ldots m_k\}, k \in \mathbb{N}\}$ of partitions $\{\mathcal{S}_{k,j}, j=1\ldots m_k\}$ of $\mathcal{S}$, with $m_{k+1} = 2m_k$, 
where $\{\mathcal{S}_{k+1,2j-1},$ $\mathcal{S}_{k+1,2j}\}$ is a partition of $\mathcal{S}_{k,j}$, with ''local parameter value'' $x = \lim_{k\rightarrow \infty}x(\mathcal{S}_{k,j})$ and function $T(x) = \lim_{k\rightarrow \infty}E_{k,j}/S_{k,j}$, $T(x)$ is a smooth function of $x$.

\subsection{Definition: Parameter region}

The parameter region of a smooth system $\mathcal{S}$ is the topological space $\mathcal{M}$ associated to (any) sequence of partitions as defined above in the limit $k\rightarrow \infty$ and with parameters $x$.

\vspace{3mm}
The function $T(x)$ is the (parameter-dependent) temperature. It follows:
\begin{enumerate}
\item $T(x)$ is uniquely defined for a smooth system and given parameterisation.
\item $\forall \epsilon \in \mathbb{R}^+$, $\exists$ partition $\{\mathcal{S}_j, j=1,\ldots,m\}$ with $|T(x)-T(y)| < \epsilon$ for any pair of parameters $x, y$ of $\mathcal{S}_j$ and $\forall j$.
\end{enumerate}

\vspace{3mm}
Smooth systems and their parameter regions are important for the development of a consistent theory, because slight departures from thermal equilibrium must be small enough and well behaved in order for differential equations of the type $f \sim \nabla T$ to hold and for the ''heat flow'' $f$ to make physical sense.

\subsection{Claim: Local trivialization of the fibre bundle over a parameter region}

Be $\mathcal{M}$ the parameter region of a smooth system. Then, $\mathcal{M}$ is a smooth manifold and one can find a locally trivial fiber bundle $E$ over $\mathcal{M}$ with standard fiber $V$, and $V$ is isomorphic to $\mathbb{R}^n$ with $n \in \mathbb{N}$.

\paragraph{Proof:}
By construction, $\mathcal{M}$ is a continuous topological space. In addition, $T(x)$ is smooth. This condition restricts the possible transformations of the parameterisation to smooth transformations, and their inverses must be smooth as well. Thus, the transformations are diffeomorphisms.
Consider a partition of the smooth system defined by $\mathcal{M}$. For each member $\mathcal{S}$ covered by one or more systems $\mathcal{S}_k:>\mathcal{S}$ according to arbitrary ordering, where $k=1, \ldots , n(\mathcal{S})$ and $n(\mathcal{S}) \in \mathbb{N}$, every $\mathcal{S}_k$ itself may be partitioned in order to generate a sequence of parameter values $x^k$ across $\mathcal{S}_k$ without ramifications, from the left-end parameter value $y_0$ (at the interface between $\mathcal{S}$ and $\mathcal{S}_k$) to the right-end parameter value $y_k$ (at the opposite end of $\mathcal{S}_k$, to which the highest order quanta are attributed). We may therefore introduce a map $\pi\big|_\mathcal{S}: \mathbb{R}^n \rightarrow U(\mathcal{S})$, where $U(\mathcal{S})$ is an open neighbourhood of $\mathcal{S}$, so that the origin is mapped to the center of $\mathcal{S}$ and the $n$ axes are scaled using the $x^k$. 

\paragraph{Subclaim} 
Any smooth system has a finite partition of systems $\mathcal{S}$ such that, given a partition $\{\mathcal{S}_j, j=1,\ldots m\}$ of $\mathcal{S}$ into subsystems $\mathcal{S}_j$, then $n(\mathcal{S}_j) = n(\mathcal{S}) \ \forall j=1,\ldots m$. \\
{\it Proof:} $T(x)$ is smooth along the map of each ($k$th) axis, $\pi(\{x^i,x^i=0 \ \forall i \ne k\})$. Also, in the sequence of partitions $\{\mathcal{S}_{k,j}, j=1\ldots m_k\}$ from the definition of $T(x)$ implying convergence, the values of $E_{k,j}/S_{k,j}$ must not change more than any given $\epsilon \in \mathbb{R^+}$ for $k>k_0$ and sufficiently large $k_0$. Thus, if $k_0$ is large enough, no sudden increase or decrease of $n(\mathcal{S}_{k,j})$ is allowed for any $k>k_0$, $n(\mathcal{S}_{k+1,2j-1})
=n(\mathcal{S}_{k,j})\pm 1$, because the sub-subsystems of one extradimension would have to be projected onto or extracted out of the lower-dimensional subsystem and thus cause the convergence of $T(\mathcal{S}_{k,j})$ to break down. Then, one can always identify $\mathcal{S}$ with one of the systems, $\mathcal{S}_{k_0,j_0}$, of the above-mentioned partition, and thus no change of $n(\mathcal{S}_{k,j})$ is possible for any member of the sequence. Thus, for any partition of $\mathcal{S}$, each subsystem $\mathcal{S_{\rm sub}}$ satisfies $n(\mathcal{S_{\rm sub}}) = n(\mathcal{S})$, as claimed. Because this statement is valid independently of the location in the parameterisation, we may use $n$ instead of $n(\mathcal{S})$ and call it the dimension of $\mathcal{M}$. \\ 

\vspace{3mm}
Let us return to our above map $\pi$. Consider again the covers $\mathcal{S}_k$ of $\mathcal{S}$. We introduce $n$ maps $v_k: C^\infty(\mathbb{R}^n) \rightarrow C^\infty(\mathbb{R}^n)$, 

\begin{equation}
\label{eq:vk}
v_k(f) = \lim_{p_k \rightarrow p} \frac{f[\pi^{-1}(p_k)] - f[\pi^{-1}(p)]}{\pi^{-1}(p_k)] - \pi^{-1}(p)},
\end{equation}

\noindent where the limit ($\lim_{p_k \rightarrow p}$) is performed by using a sequence of partitions. One can use the above sequence of parameter values $x^k$ across $\mathcal{S}_k$ without ramifications to verify that $v_k(f)$ is linear in $f\in C^\infty(\mathbb{R}^n)$ and satisfies the Leibniz rule $v_k(fg) = v_k(f)g+fv_k(g)$. Thus, the $v_k$ define a vector space $V_p$ isomorphic to $\mathbb{R}^n$, which can be attached to each point $p$ with parameter $x$ of the parameter region.

Moreover, $\mathcal{M}$ may be covered with open sets $\mathcal{O}_\alpha$ with non-empty intersections, the maps \mbox{$\pi^{-1}_\alpha: \mathcal{O}_\alpha \rightarrow V_p, p \in O_\alpha$} are one-to-one and onto with transition functions $\Phi_\beta\circ\Phi_\alpha^{-1}$ defined on the intersections. 
The functions $\Phi_\beta\circ\Phi_\alpha^{-1}$ define the transformations and are therefore diffeomorphisms. Thus, $\mathcal{M}$  is a smooth manifold. 
This structure also defines a fiber bundle $\pi: E \rightarrow \mathcal{M}$. Every map $V_p \rightarrow U(p)$ to a neighbourhood $U(p)$ of $p$ defines a fibre. 

Furthermore, all $V_p$ are isomorphic to each other. Thus, by analogy to the parallel transport of tangent vectors, we may always find a neighbourhood $U(p)$ small enough in order to transport any vector from $p$ to $q \in U(p)$, $\forall p \in \mathcal{M}$. Thus, we can make all the fibers $\pi\big|_q$, $q \in U(p)$ identical, {\rm i. e.} trivialize $E \big|_U$. Therefore, $\pi: E \rightarrow \mathcal{M}$ has a local trivialization with (locally defined) standard fiber $F$, $\Phi: E \big|_U \rightarrow U \times F$, and $F$ is isomorphic to $\mathbb{R}^n$. This completes the proof of the above claim on the local trivialization of the fibre bundle over a parameter region.

\subsection{Definition}
A space-time region $\mathcal{M}$ is called thermally small if 

\begin{equation}
\label{eq:small_M}
\frac{[T^{-1}(p)-T^{-1}(q)] \int_\mathcal{M}}{\int_\mathcal{M} |T^{-1}| } \ll 1
\end{equation} 

\noindent for any two points $p, q \in \mathcal{M}$. 

%%%%%%%%%%%%% Appendix B

\section{}
\label{sec:B}

In this appendix, it shall be shown that the dimension $n = 3+1$ of a macroscopic parameterisation manifold maximises the ratio of locally resulting degrees of freedom per imposed degree of freedom of the theory.

\subsection{Imposed degrees of freedom}

Consider a macroscopic parameterisation manifold of dimension $n>1$ and a locally trivial fiber bundle with normal fiber $F$ given by a vector space $V$ isomorphic to $\mathbb{R}^n$. From an arbitrary location, any other nearby location can be reached by a displacement along an arbitrary direction of the parameterisation (as defined by specifying $n$ degrees of freedom) and by a change of direction at an arbitrary distance (or local rotation, as uniquely defined by a rotation matrix, {\rm i. e.} a matrix specified by $n^2$ degrees of freedom)\footnote{Although there is a constraint on the determinant of a rotation matrix, one requires all the matrix elements in order to discribe the rotation.}. Displacements and rotations are the two basic tools and normally impose $n(n+1)$ degrees of freedom. However, if $n<4$, the rotation matrix may be replaced by an axial vector specified by $n$ degrees of freedom. In this case, only $2n$ degrees of freedom are imposed.

\vspace{3mm}
It is possible to break the explicit $n$-fold symmetry of the vector space by Wick-rotating one dimension ($x^n \rightarrow i\cdot x^n$), whence general ''rotations'' may be uniquely defined by decomposing them into rotations in the $n-1$-dimensional Euclidean subspace and Lorentz boosts between the $n$th dimension and each Euclidean dimension. Thus, the boosts specify $n-1$ degrees of freedom. For $n=3+1$, we have thus $4+3+3=10$ degrees of freedom (4 translations, 3 rotations and 3 boosts).

\subsection{Locally resulting degrees of freedom}

All resulting phenomena must depend on the thermodynamic properties of the parameterisation space, {\rm i. e.} the transformation vector fields which specify the space configuration, given by $n$ basis vectors of dimension $n$. This yields a total of $n^2$ locally resulting degrees of freedom\footnote{No constraints on the transformation vector fields are taken into account at this stage, because any constraint may still depend on the outcome of thermodynamic computations.}.

\subsection{Ratio of locally resulting per imposed degree of freedom}

From the above consideration, we find that the maximum ratio is obtained for $n=3+1$. We obtain a ratio of $4^2/[4+3+3]=16/10$.
 
\vspace{3mm}
 One could suggest to use more than one Wick-rotated dimension. It shall be shown why this does not lead to further optimisation.
 
\vspace{3mm}
Consider two or more Wick-rotated dimensions, leading to more than one pure imaginary coordinate. Such a space could be back-Wick-rotated to yield the extended Minkowski metric 

\begin{equation}
(\eta_{\mu\nu}) = diag(+1,+1,+1,\ldots,-1,-1,-1).
\end{equation}

\noindent However, it would then be possible to move smoothly away from this metric by diffeomorphisms, $\eta_{\mu\nu} \rightarrow g_{\mu\nu} = \eta_{\mu\nu} + \epsilon_{\mu\nu} \rightarrow {g'}_{\mu\nu}$ and so on, preserving the sign of $det(g_{\mu\nu})$, and to finally end up with $g^f_{\mu\nu}$ in the form:

\begin{equation}
(g^f_{\mu\nu}) = diag(-1,-1,+1,\ldots,-1,-1,-1),
\end{equation}

\noindent which still preserves the sign of the determinant. In this way, the Euclidean subspace could be enlarged, and one would have to introduce rotation matrices of higher Euclidean subdimension for the imposed degrees of freedom, and no optimisation would be possible, as claimed.

\vspace{3mm}
Another possibility is to use distinguishable ''imaginary numbers'' for Wick-rotated coordinates, which is mathematically equivalent to introducing partly or fully quaternion-valued coordinates, {\rm e. g.}

\begin{equation}
x^1 \rightarrow i\cdot x^1, \quad x^2 \rightarrow j\cdot x^2, \quad x^3 \rightarrow k\cdot x^3, \quad x^l \rightarrow x^l, \quad (l=4,\ldots 6).
\end{equation}

\noindent However, the space $V$ would have to be based on quaternions which have non-commutative product. Therefore, distinguishable ''imaginary numbers'' have to be excluded. We can see it also in another way: The extended Minkowski metric of such a space would read

\begin{equation}
(\eta_{\mu\nu}) = diag(+1+,1,+1,-1,-1,-1),
\end{equation}

\noindent and this metric could not be distinguished from the one corresponding to the multiple imaginary structure

\begin{equation}
x^1 \rightarrow i\cdot x^1, \quad x^2 \rightarrow i\cdot x^2, \quad x^3 \rightarrow i\cdot x^3, \quad x^l \rightarrow x^l, \quad (l=4,\ldots 6).
\end{equation}

\noindent Thus, for a space containig purely bosonic matter (without torsion) \cite{Mandrin2}, Einstein notation (using the metric and the Levi-Civita connection) would be fully sufficient, and no distinction would be possible between quaternion-valued and purely imaginary Wick-rotated coordinates. Therefore, we would have to reintroduce rotation matrices of dimension $n^2$ for the imposed degrees of freedom, and no optimisation would be possible.

%%%%%%%%%%%%% Appendix C

\section{}
\label{sec:C}

In this appendix, the derivation of the entropy of the gravitational field without interactions is summarised \cite{Mandrin2}. We start with Eqn. (\ref{eq:var_boundary_tau}):

\begin{equation}
\label{eq:C_var_boundary_tau}
\delta S\bigg|_{\partial\mathcal{M}} = \sum_{a} \int_{\partial\mathcal{M}_a} {\rm d}^3x \ \sqrt{|\gamma|} \ e^I_i \delta \tau^i_{a I}.
\end{equation}

The boundary term $\delta S\bigg|_{\partial\mathcal{M}}$ can be expressed in terms of the flow of $\delta \tau^i_{a I}$ across the boundary \cite{Mandrin2}: $\delta\tau^i_{a I} = \delta\tau^{i\gamma}_{a I} n_\gamma$, where $n_\gamma$ is the unit vector normal to the subspace associated to the boundary $\partial\mathcal{M}_a$.
We may thus apply Gauss' law on Eqn. (\ref{eq:C_var_boundary_tau}) to obtain

\begin{equation}
\label{eq:gauss}
\sum_a \int_{\partial\mathcal{M}_a} {\rm d}^3x \sqrt{|\gamma|} e^I_i \delta \tau_{a I}^{i\gamma} n_\gamma = \int_\mathcal{M} d^4x \sqrt{-g} \nabla_\gamma(e^I_\alpha \delta \tau_{a I}^{\alpha\gamma} )
\end{equation}

Using Leibnitz' Rule, we obtain \cite{Mandrin2}:

\begin{equation}
\label{eq:dUpsilon}
\label{eq:var_Upsilon}
\nabla_\gamma(e^I_\alpha \delta \tau_{a I}^{\alpha\gamma} ) = e^I_\mu \ e^J_\nu \ \delta\Phi^{\mu\nu}_{IJ} + \omega_{\mu IJ} \ \delta \Omega^{\mu IJ},
\end{equation}

\noindent where 

\begin{eqnarray}
\label{eq:dPhi}
\delta\Phi^{\alpha\delta}_{IJ} & = & \eta_{IJ} g^{\delta\epsilon} \nabla_\gamma [e^K_\epsilon \delta \tau_{aK}^{\alpha\gamma}], \\
\label{eq:dOmega}
\delta\Omega^{\mu IJ} & = & (e^J_\alpha \eta^{KI} + e^I_\alpha \eta^{KJ}) \delta \tau_{aK}^{\alpha\mu}, \\
\label{eq:omega}
\omega_{\mu IJ}  & = & e^\alpha_I \ \nabla_\mu e_{\alpha J}.
\end{eqnarray}

Eqns (\ref{eq:gauss}) and (\ref{eq:var_Upsilon}), the boundary contribution to the entropy variation, and the bulk contribution to the entropy variation must be complementary and, together, yield the total entropy, which is fully analogue to the procedure for the action variation in \cite{Mandrin1} or according to \cite{York}:

\begin{eqnarray}
\delta S_{\rm total} & = & \int_\mathcal{M} d^4x \ [\sqrt{-g} (e^I_\mu \ e^J_\nu \ \delta \Phi^{\mu\nu}_{IJ} + \omega_{\mu IJ} \ \delta \Omega^{\mu IJ}) \nonumber \\
\label{eq:C_delta_Total_grav}
& + &  \Phi^{\mu\nu}_{IJ} \ \delta (\sqrt{-g} \ e^I_\mu \ e^J_\nu) + \Omega^{\mu IJ} \ \delta (\sqrt{-g} \ \omega_{\mu IJ})], \\
\label{eq:C_Total_grav}
S_{\rm total} & = & \int_\mathcal{M} d^4x \sqrt{-g} \ [e^I_\mu \ e^J_\nu \ \Phi^{\mu\nu}_{IJ} + \omega_{\mu IJ} \ \Omega^{\mu IJ}].
\end{eqnarray}

%%%%%%%%%%%  References



%\begin{thebibliography}{10}

\end{document}